\documentclass{article}
\usepackage{ijcai24}

\usepackage{times}
\usepackage{soul}
\usepackage{url}
\usepackage[hidelinks]{hyperref}
\usepackage[utf8]{inputenc}
\usepackage[small]{caption}
\usepackage{graphicx}
\usepackage{amsmath}
\usepackage{amsthm}
\usepackage{booktabs}
\usepackage{algorithm}
\usepackage{algorithmic}
\usepackage[switch]{lineno}
\usepackage{multirow}
\usepackage{multicol}
\usepackage{subfigure}
\usepackage{subcaption}
\usepackage{amsfonts}
\usepackage{bm}
\usepackage{verbatim}
\usepackage{makecell}
\usepackage{xcolor}

\title{Arrange, Inpaint, and Refine: Steerable Long-term Music Audio Generation and Editing via Content-based Controls}

\author{
Liwei Lin$^{1}$
\and
Gus Xia$^{2,1}$\and
Yixiao Zhang$^3$\And
Junyan Jiang$^{1}$\\
\affiliations
$^1$Music X Lab, New York University Shanghai\\
$^2$Music X Lab, Mohamed bin Zayed University of Artificial Intelligence\\
$^3$C4DM, Queen Mary University of London
\emails
liwei.lin@nyu.edu,
Gus.Xia@mbzuai.ac.ae,
yixiao.zhang@qmul.ac.uk,
jj2731@nyu.edu
}

\begin{document}

\maketitle

\begin{abstract}

Controllable music generation plays a vital role in human-AI music co-creation. While Large Language Models (LLMs) have shown promise in generating high-quality music, their focus on autoregressive generation limits their utility in music editing tasks. To address this gap, we propose a novel approach leveraging a parameter-efficient heterogeneous adapter combined with a masking training scheme. This approach enables autoregressive language models to seamlessly address music inpainting tasks. Additionally, our method integrates frame-level content-based controls, facilitating track-conditioned music refinement and score-conditioned music arrangement. We apply this method to fine-tune MusicGen, a leading autoregressive music generation model. Our experiments demonstrate promising results across multiple music editing tasks, offering more flexible controls for future AI-driven music editing tools. The source codes and a demo page showcasing our work are available at https://kikyo-16.github.io/AIR.

\end{abstract}

\section{Introduction}
In recent years, there has been a surge in the development of deep music generative models, encompassing both audio and symbolic domains~\cite{huang2018music,huang2020pop,dhariwal2020jukebox,wang2021musebert,zhao2021accomontage,copet2023simple,agostinelli2023musiclm}. Particularly, the emergence of Large Language Models (LLMs) has enabled various systems to achieve high-quality generations. Such progress offers a glimpse into the potential of constructing effective tools for facilitating human-AI co-creation. 

Current LLM-based music audio generation models exhibit remarkable performance~\cite{dhariwal2020jukebox,agostinelli2023musiclm,copet2023simple}. Some initiatives~\cite{lin2024contentbased} aim to equip these models with content-based controls, such as drum patterns and chord progressions, showcasing their potential for flexible music editing tasks. However, as these models focus on autoregressive generation, they lack the capability of inpainting, a common practice in music creation involving iterative refinements of arbitrary audio segments. Existing music inpainting models~\cite{zhou2019vision,marafioti2020gacela,garcia2023vampnet}, while attempting to address this need, struggle with controllable inpainting and capturing long-term music context dependencies. Consequently, they are primarily effective for short-range gap filling, typically limited to one or two seconds, and lack controllability.

To this end, we introduce AIRGen, a model capable of
doing \textbf{a}rrangement, \textbf{i}npainting, and \textbf{r}efinement by applying a novel parameter-efficient heterogeneous adapter to fine-tune MusicGen~\cite{copet2023simple}, an advanced autoregressive music audio language model. Our heterogeneous adapter re-purposes the autoregressive language model to seamlessly solve mask-filling tasks, thereby equipping MusicGen with music in-painting ability (Fig.~\ref{fig:editingA}). Moreover, the heterogeneous adapter integrates frame-level audio controls into the model architecture, enabling precise content-based manipulation, including drum tracks, chord progression, and piano covers. This enables track-conditioned music refinement (Fig.~\ref{fig:editingB}) and score-conditioned music arrangement (Fig.~\ref{fig:editingC}). 

In summary, the contributions of this work are as follows:

\begin{figure*}[t]
  \centering
  \subfigure[]{\label{fig1a}\includegraphics[width=.33\textwidth, page=10, clip, trim=2.4cm 3cm 5.1cm 1.5cm]{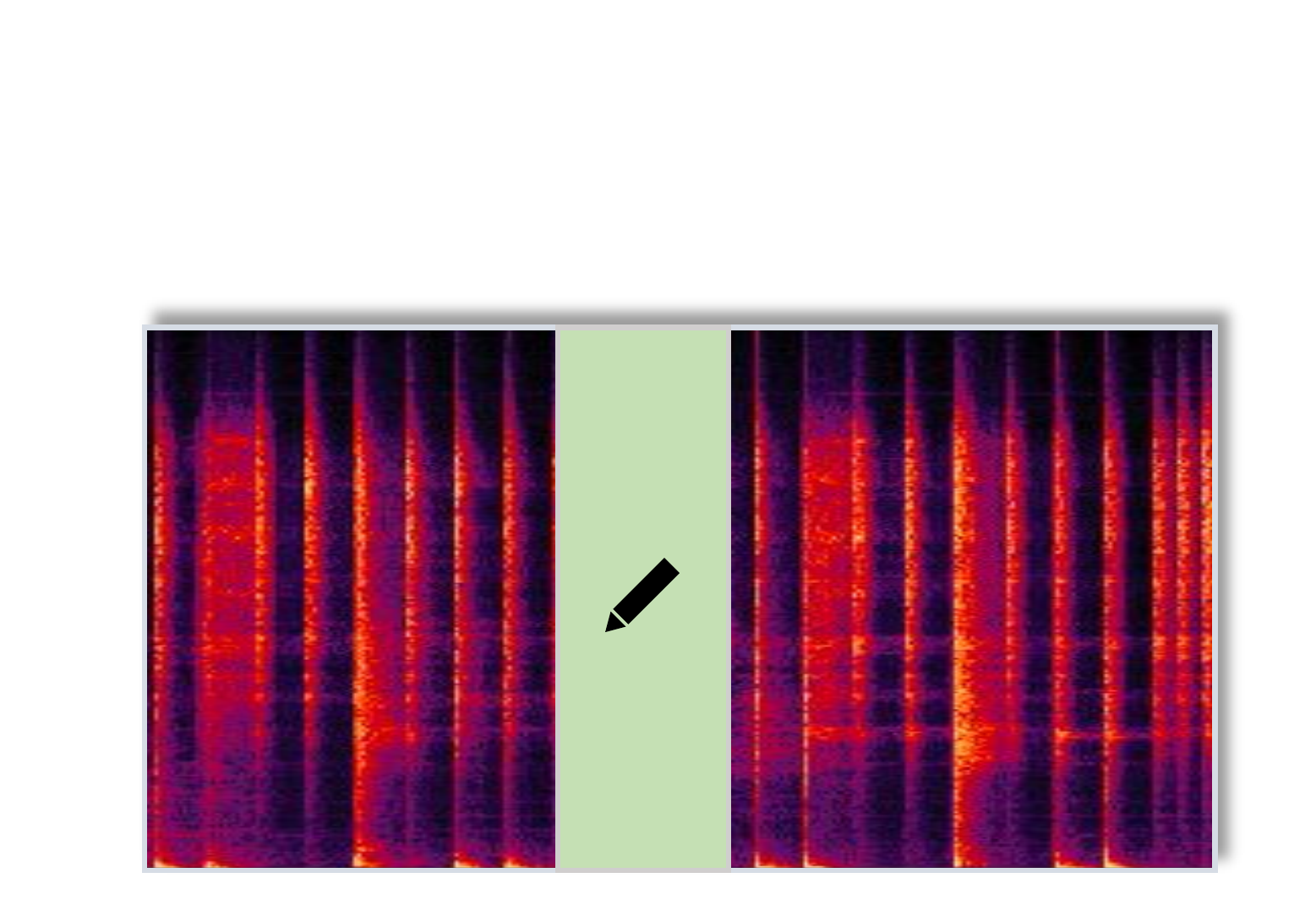}\label{fig:editingA}}
  \subfigure[]{\label{fig1b}\includegraphics[width=.33\textwidth, page=11, clip, trim=2.4cm 3cm 5.1cm 1.5cm]{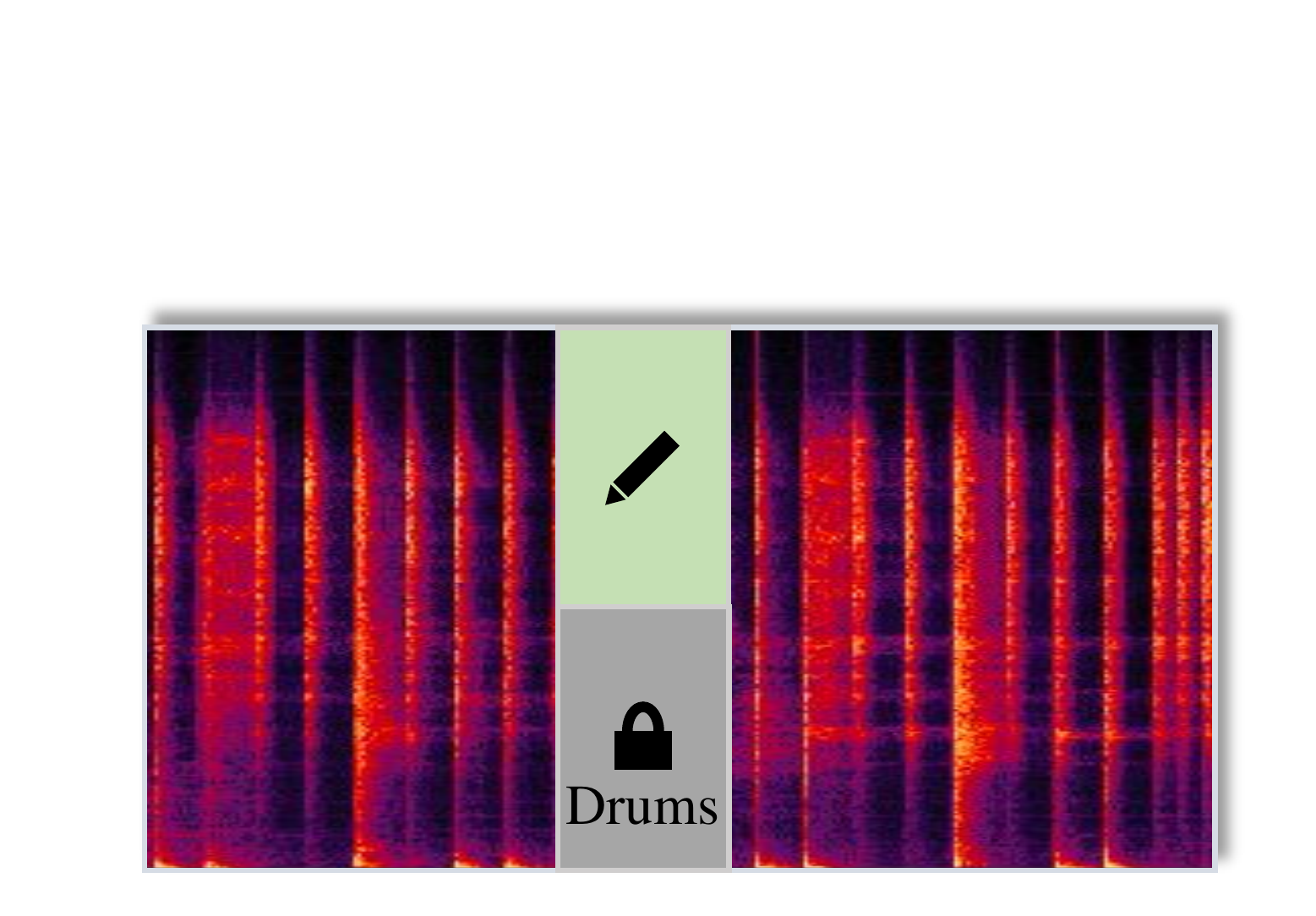}\label{fig:editingB}}
  \subfigure[]{\label{fig1c}\includegraphics[width=.33\textwidth, page=12, clip, trim=2.4cm 3cm 5.1cm 1.5cm]{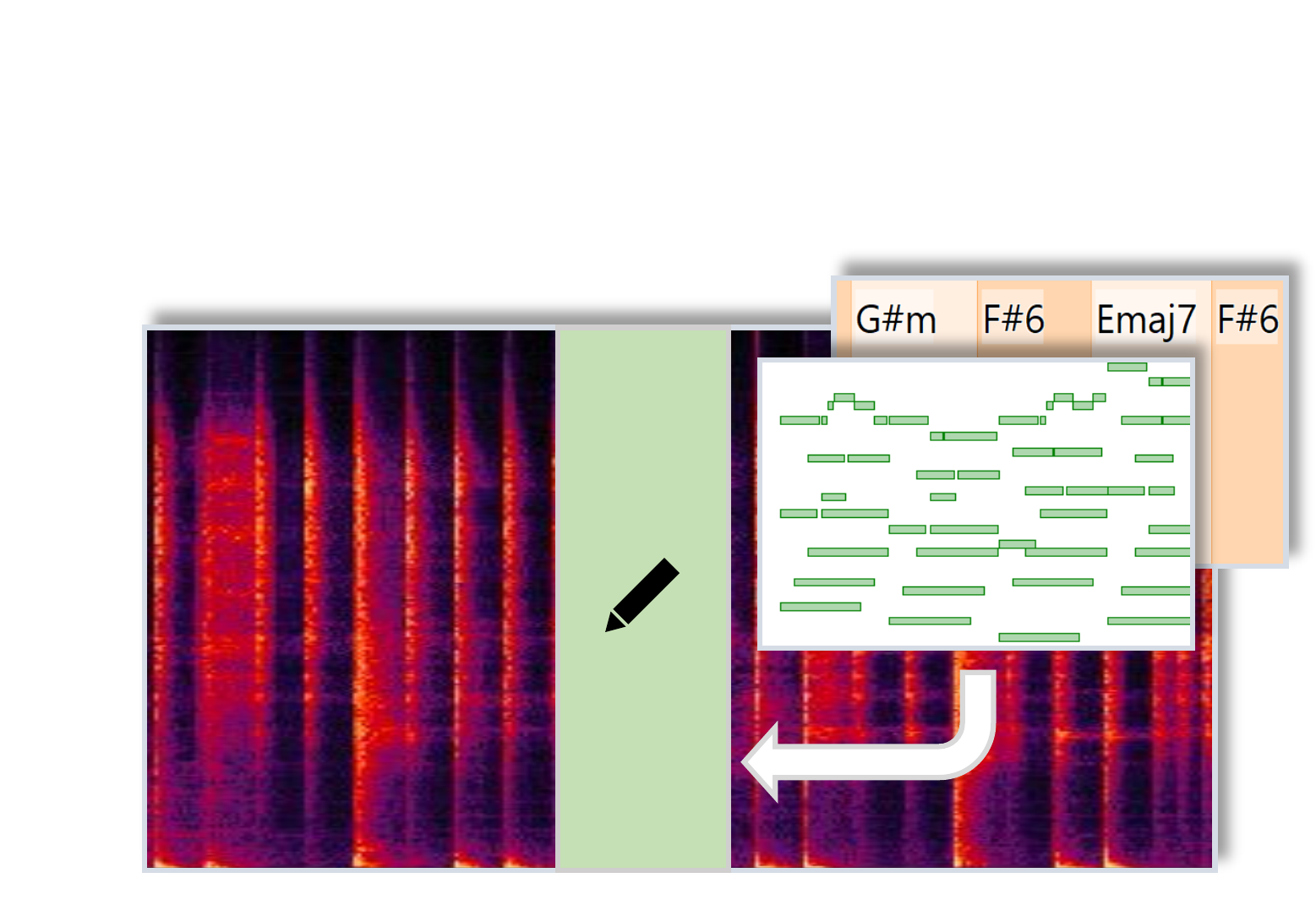}\label{fig:editingC}}
  \caption{Different music editing tasks accomplished by the model. Light green blocks with pen icons denote the masked parts to generate. (a)~To inpaint an entire segment; (b)~To refine some tracks conditioned on other tracks (e.g., drums); (c)~To arrange the target segment following user-provided piano cover or chord controls.}
  \label{fig:editing}
\end{figure*}

\textbf{Parameter-efficient heterogeneous adapter.} We introduce a novel heterogeneous adapter, combined with a well-designed masking training scheme. This approach successfully transforms a pretrained autoregressive LM into a masked LM, while incorporating sequential controls through Parameter-Efficient Fine-Tuning.

\textbf{Flexible long-term music editing.} As shown in Figure~\ref{fig:editing}, our model enables flexible editing of arbitrary segments of the audio mixture. This includes re-generating (inpainting) the segment based on its past and future contexts, refining the segment based on prescribed audio content (e.g., the drum track), and re-arranging the segment given semantic controls such as chord progressions and piano arrangements. Additionally, our model is adept at filling segments lasting more than 8 seconds.

\textbf{High-quality steerable Generation.} Experimental results demonstrate that our model outperforms existing baselines at the inpainting task. Additionally, for the refinement and arrangement tasks, our model achieves comparable quality to unconditioned autoregressive generation, a capability not observed in other models.

\section{Related Work}
\subsection{General Audio and Music Generation Models}
Music audio generation necessitates extensive contextual modeling to account for the intricate structure of musical language. Recent large-scale music audio generative models, encompassing autoregressive and diffusion-based approaches, have made remarkable strides in capturing such a long-term structure while introducing cross-modal conditions. For example, Jukebox~\cite{dhariwal2020jukebox} leverages VQ-VAE~\cite{van2017neural} and transformer decoders to achieve lyric and genre-based generation; Diffusion-based Moûsai~\cite{schneider2023mo} adopts the pretrained frozen T5 encoder~\cite{raffel2020exploring} to summarize text conditions; autoregressive MusicGen~\cite{copet2023simple} realizes monophonic melody and text controls by assembling EnCodec~\cite{defossez2022high}, T5 encoder, and an acoustic transformer decoder. Additionally, Coco-Mulla~\cite{lin2024contentbased} and Music ControlNet~\cite{wu2023music} facilitate more flexible music generation by incorporating content-based controls.

\subsection{Music Inpainting}
Music inpainting models support music generation in more diverse and interactive scenarios. In the symbolic music domain, the inpainting duration is usually long~\cite{ippolito2018infilling,chen2020music,chang2021variable,min2023polyffusion,10096446}. In the music audio domain, Perraudin et al.~\shortcite{perraudin2018inpainting} use a similarity graph to retrieve suitable segments in the source audio to patch up the masked section; Marafioti et al.~\shortcite{marafioti2019context} use a conv-deconv neural net to predict the masked spectrogram given the past and future contexts; Zhou et al.~\shortcite{zhou2019vision} condition the inpainting on the music performance video via a visual-audio joint feature space; GACELA~\cite{marafioti2020gacela} uses a generator with a latent condition code as input to address the multi-modality of music content; Vampnet~\cite{garcia2023vampnet} adopts that methodology and uses masked acoustic token modeling to implement music inpainting. However, in all the above-mentioned works, the inpainting duration is limited to under 2 seconds, and the models fail to meet flexible content-based controls.

\subsection{Parameter-Efficient Fine-tuning}
Parameter-efficient fine-tuning (PEFT) methods adapt pretrained language models (PLMs) by selectively updating model parameters, leading to notable reductions in computational and storage costs. Methods such as appending task-specific prefixes to input sequences~\cite{li2021prefix,liu2022p} and employing low-rank adaption (LoRA) for fine-tuning pretrained linear matrices~\cite{hu2021lora} have shown promise. Additionally, approaches like LLaMA-Adapter adjust attention outputs using prompt adapters and zero gate scalars, with a multi-modal conditional variant incorporating global image representations~\cite{zhang2023llama,gao2023llama}. Inspired by LLaMA-Adapter, we introduce a novel PEFT method in this study, designed to fine-tune large-scale models while accommodating external sequential conditions, transforming a pretrained autoregressive language model into a masked LM.

\section{Methodology}
In this section, we describe in detail the base LLM to fine-tune, the well-designed content-based controls, the tokenization scheme for input into the LLM, and the proposed heterogeneous adapter.

\subsection{MusicGen} \label{subsec:music-gen}
We take the text-only version of MusicGen as the base LLM. MusicGen is an autoregressive transformer-based model, conditioned on text or melody, using quantized tokens from an EnCodec~\cite{encodec} audio tokenizer for high-fidelity music generation. It incorporates Residual Vector Quantization to handle multiple streams of discrete tokens from learned codebooks. 

This version of MusicGen comprises three key components: a pretrained EnCodec, a pretrained T5 encoder, and an acoustic transformer decoder. The acoustic transformer decoder is composed of $N$ layers, each featuring a causal self-attention block and a cross-attention block tailored to handle conditioning text prompts. Our PEFT training process solely modifies the self-attention blocks within the acoustic transformer decoder.

\begin{algorithm}[tb]
    \caption{Piano reduction algorithm}
    \label{alg:piano-reduction}
    \textbf{Input}: $M=$ Original MIDI tracks\\
    \textbf{Output}: $P=$ Reduced piano track
    \begin{algorithmic}[1] 
        \STATE Let $\Gamma =$ Piano-like MIDI tracks in $M$.
        \STATE ${\Gamma}^{'}$ = Sort all note events in $\Gamma$ by duration in a descending order.
        \STATE Initiate an empty piano track $P$.
        \FORALL{ Note event $e$ in ${\Gamma}^{'}$}
        \STATE $e^{'} \leftarrow$ Set program of $e$ to $0$.
        \IF{$e^{'}$ does not overlap with any event in $P$}
        \STATE Add $e^{'}$ to $P$.
        \ENDIF 
        \ENDFOR
        \STATE \textbf{return} $P$
    \end{algorithmic}
\end{algorithm}

\subsection{Content-Based Controls} \label{subsec:content-based-controls}
We categorize content-based controls into two types: external and internal controls. Internal controls involve surface-level manipulation of acoustic tracks, requiring the generated audio to include the condition tracks. External controls, on the other hand, entail conceptual adjustments such as chord progression or piano cover. The apparent challenge is how MusicGen, an audio model, can process the symbolic conditions (e.g., notes, chords) prescribed by the external controls. Our solution is to algorithmically render all symbolic controls into the audio format before inputting them to MusicGen. 

In our experiments that demonstrate internal controls, we use drum tracks as the condition. 
The external controls in our experiments include chord progression and piano cover.  During audio synthesis, we render the current chord annotation into a block chord at each beat onset. For piano cover controls, we introduce a rule-based piano reduction algorithm outlined in Algorithm~\ref{alg:piano-reduction} to extract piano covers from provided MIDI annotations. Subsequently, we render the piano reduction to audio.

\subsection{Tokenization Design} \label{subsec:token-design}

We extract discrete tokens from both the target and the condition audio sequences using Encodec, operating at a resolution of 50 frames per second. Let $\bm{X}=\{\bm{x}_1, \bm{x}_2,...,\bm{x}_T\}$ $(\bm{X}\in\mathbb{R}^{T\times n})$ and $\bm{C}=\{\bm{c}_1, \bm{c}_2,...,\bm{c}_T\}$ $(\bm{C}\in\mathbb{R}^{T\times n})$ represent the extracted tokens of the target and condition sequences, respectively, where $T$ denotes the total number of frames and $n$ denotes the dimension of a discrete token. Our goal is to reconstruct the original $\bm{X}$ from a corrupted $\bm{X}$ using the condition $\bm{C}$.

\begin{figure}[t]
  \centering{
  \includegraphics[width=0.7\columnwidth]{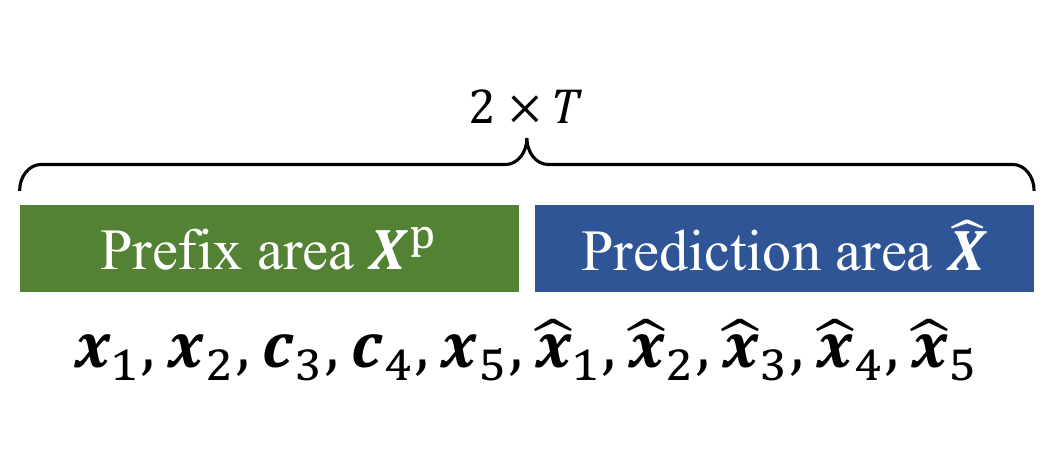}
  }
  
  \caption{Design of the input sequence, with $T=5$ as an example. The prefix area is followed by the prediction area. The prefix area spans from $1$ to $T$. Here the unmasked locations contain the infilling contexts. At the masked locations, the ground-truth audio is masked away and replaced by the frame-level condition audio track. The prediction area spans from $T+1$ to $2 \times T$. In this example, the masks happen to be contiguous, while in general an arbitrary set of frames can be masked.}
  \label{fig:sequence_design}
\end{figure}
We prepend the combination of $\bm{C}$ and corrupted $\bm{X}$ as the prefix of the input sequence to the decoder (Figure~\ref{fig:sequence_design}). Denote the prefix as $\bm{X}^{\rm p}=\{\bm{x}^{\rm p}_1, \bm{x}^{\rm p}_2, ..., \bm{x}^{\rm p}_T\}$ $(\bm{X}^{\rm p}\in\mathbb{R}^{T\times n})$, 
\begin{equation}
\bm{x}^{\rm p}_t=\begin{cases}\mathbf{c}_t, & \text{
if $t$-th frame is masked,
}\\ 
\bm{x}_t, & \text{
otherwise.
} 
    \end{cases}
\end{equation}
Given $\hat{\bm{x}}_1,\hat{\bm{x}}_2, ..., \hat{\bm{x}}_t$,  which are the predicted tokens up to $t$, then the next predicted audio token $\hat{\bm{x}}_{t+1}\in\mathbb{R}^n$ is,
\begin{equation}
    \hat{\bm{x}}_{t+1}={\rm Decoder}([\bm{X}^{\rm p}; \hat{\bm{x}}_1,\hat{\bm{x}}_2, ..., \hat{\bm{x}}_t]).
\end{equation}

\begin{figure*}[t]
  \centering  \includegraphics[width=2\columnwidth]{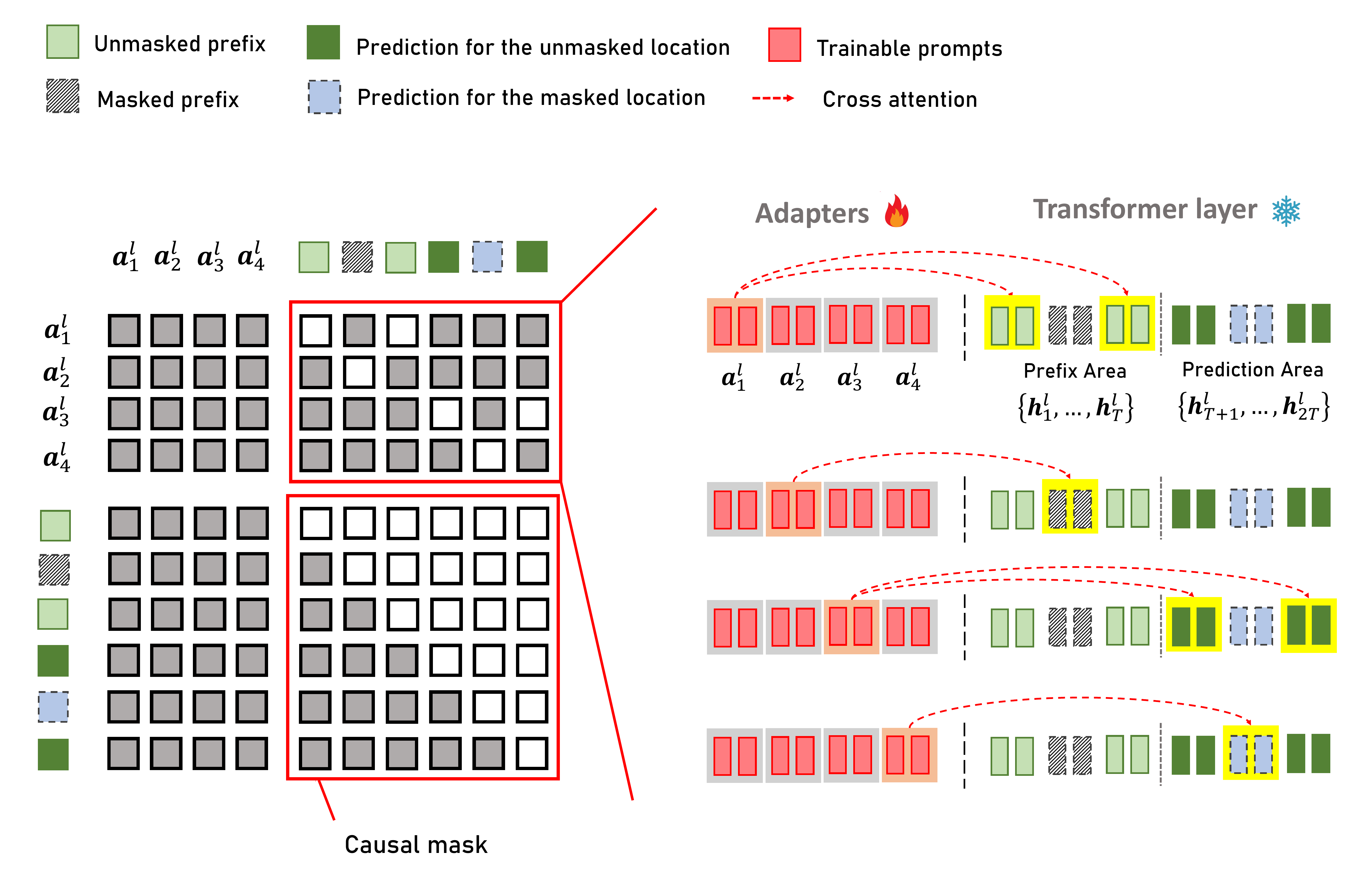}
  \caption{The attention mask in heterogeneous adapters. The top-right block in the attention mask matrix indicates the cross-attention between learnable adapters and input tokens, where a specific group of adapters associates with a specific type of input tokens. The bottom-right block of the attention mask matrix is a general causal mask, regardless of the distribution of the masked tokens in the prefix/prediction area. In this example, the masked tokens are contiguous; however, in general an arbitrary set of frames can be masked.}
  \label{fig:adapters}
\end{figure*}

During training, we apply cross-entropy loss to the prediction area for both masked and unmasked tokens,

\begin{equation}
    L=\frac{1}{T}\sum_{t=1}^{T}{\rm CrossEntropy}(\bm{x}_t, \hat{\bm{x}}_{t}).
\end{equation}
This design enables the autoregressive decoder to capture full-context information and incorporate frame-wise condition information when inpainting corrupted audio.

\subsection{Heterogeneous Adapter} \label{subsec:hete-adapter}

To enhance decoder fine-tuning efficiency, we introduce a heterogeneous adapter, where multiple different lightweight adapters are trained to work together in the same decoder. 

In vanilla PEFT with adapters, the same type of adapters is applied to all frames since the signal $\bm{X}^{\rm p}$ is homogeneous in a typical LM task. In our method, each layer incorporates four different types of trainable adapters, with each type directed at frames of a specific type, as depicted in Figure~\ref{fig:adapters}. Let ${\bm a}^l_1, {\bm a}^l_2, {\bm a}^l_3, {\bm a}^l_4 \in \mathbb{R}^{m\times d}$ represent the adapters of the $l$-th self-attention layer, where $m$ is the number of adapter vectors per type and $d$ is the dimension of both the decoder's hidden state and each adapter vector. We deploy ${\bm a}^l_1$ and ${\bm a}^l_3$ to incorporate the infilling context at the unmasked locations and deploy ${\bm a}^l_2$ and ${\bm a}^l_4$ to take in condition information and perform inpainting at the masked locations. 

Let $\bm{H}^{l}=\{\bm{h}^{l}_1,\bm{h}^{l}_2,...,\bm{h}^{l}_{2T}\}$ denote the input representation of the $l$-th self-attention layer. The heterogeneous adapter modifies the attention output of $\bm{h}^{l}_t$ by computing cross-attention between $\bm{h}^{l}_t$ and $\bm{a}^{l}_{r(t)}$. The type $r(t)$ is selected based on whether the $t$-th frame is in the prefix area or the prediction area and whether it is masked:
\begin{equation}
r(t)\left\{\begin{matrix}
1,& \text{if $t\leq  
 T$ and $t$-th frame is unmasked},\\ 
2,& \text{if $t\leq  
 T$ and $t$-th frame is masked},\\ 
3,& \text{if $t>T$ and $(t-T)$-th frame is unmasked},\\ 
4,& \text{otherwise.}
\end{matrix}\right.
\end{equation}

Before being modified by heterogeneous adapters, each layer of the transformer decoder computes self-attention $\bm{S}^l=\{\bm{s}_1^l, \bm{s}_2^l,...\bm{s}_{2T}^l\}$ among the hidden states with a causal mask:

\begin{equation}
    \bm{S}^l  = \text{Self-Attention}({\bm H}^l).\label{eq:self-attn}
\end{equation}

The heterogeneous adapters introduce a cross-attention mechanism, between learnable adapters and the hidden states of the transformer layers, as shown in Figure~\ref{fig:adapters}. This cross-attention reuses all the projection matrices of the original transformer decoder, which are used to perform self-attention as described in Equation~\ref{eq:self-attn}. The cross-attention $\bm{u}^{l}_t$ for $\bm{h}^{l}_t$ is computed as follows:

\begin{equation}
\bm{u}^{l}_t=\text{Cross-Attention}(\bm{h}_t^l,\bm{a}^l_{r(t)}).
\end{equation}


After being modified by the adapters, the adjusted attention output $\bm{s}^{l,*}_t$ is given by:

\begin{equation}
\bm{s}^{l,*}_t=\bm{s}^{l}_t + g^{l}_{r(t)}\cdot\bm{u}^{l}_t,
\end{equation}
where $g^{l}_{r(t)}$ is a learnable gating factor initialized to zero.


Throughout the training, all parameters in the decoder remain frozen except for the adapters $\{{\bm a}^l_1, {\bm a}^l_2, {\bm a}^l_3, {\bm a}^l_4\}$ and the gates $\{g^l_1, g^l_2, g^l_3, g^l_4\}$ at each layer.

\section{Experiments}
\subsection{Dataset}
Our model is trained on the training split of Slakh2100, a dataset of synthesized MIDI files using a high-quality soundfont \cite{manilow2019cutting}. We derive pseudo chord and beat annotations using Madmom \cite{bock2016madmom} and a chord recognition model \cite{jiang2019large}. The total training set comprises 1289 instrumental songs with silence trimmed away at both ends. Evaluation is conducted on both the test split (151 songs) of Slakh2100 and RWC-POP100, a dataset featuring 100 real-recording pop songs with MIDI, chord, and beat annotations \cite{goto2002rwc}.

\subsection{Training}
We train three models: Drum-AIR, Chord-AIR, and Piano-AIR, tailored for drum-track conditioning, chord-progression conditioning, and piano-cover conditioning, respectively.

Each model is trained using two A800\rq s with an initial learning rate of 2e-3, a batch size of 24, and a sample duration of 15 seconds, for 10 epochs, totaling 8 hours of training time. The warm-up epoch is set to 1, and the model is updated using a cross-entropy loss.

\subsection{Evaluation}
\begin{figure}[t]
\centering
{
    \subfigure[Mask Pattern~1.]{
        \includegraphics[width=0.3\columnwidth,page=1]{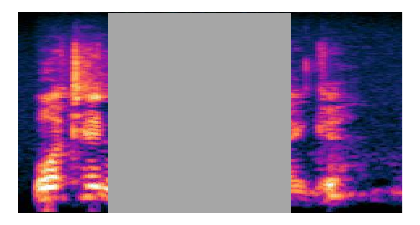}
    }
    \subfigure[Mask Pattern~2.]{
        \includegraphics[width=0.3\columnwidth,page=2]{figs/mask_123.pdf}
    }
    \subfigure[Mask Pattern~3.]{
        \includegraphics[width=0.3\columnwidth,page=3]{figs/mask_123.pdf}
    }
}
  \caption{
  The three types of masks used in evaluation. The number of masks is by design and the location is random.
}
  \label{fig:mask-123}
\end{figure}

\begin{table}[t]
\renewcommand{\arraystretch}{1.3}
\setlength{\tabcolsep}{3mm}
\centering
\subtable[Slakh2100 Test Set]{
\begin{tabular}{ c|cc|cc }
    \Xhline{3\arrayrulewidth}
    &\multicolumn{2}{c|}{\textbf{CLAP}$_\text{src}\uparrow$}&\multicolumn{2}{c}{\textbf{FAD}$_\text{vgg}\downarrow$}
    \\
    \cline{2-5}
    &Full&Prefix&Full&Prefix
    \\
    \Xhline{3\arrayrulewidth}
    Drum-AIR&0.749&0.756&1.423&1.422\\
    Chord-AIR&0.753&0.757&\textbf{1.220}&1.222\\
    Piano-AIR&\textbf{0.755}&\textbf{0.761}&1.290&1.282\\
    \Xhline{2\arrayrulewidth}  
    MusicGen&0.656&0.687&1.251&\textbf{1.218}\\
    VampNet&0.631&0.643&2.910&3.424\\
    \Xhline{3\arrayrulewidth}
\end{tabular}
}
\subtable[RWC-POP-100]{
\begin{tabular}{ c|cc|cc }
    \Xhline{3\arrayrulewidth}
    &\multicolumn{2}{c|}{\textbf{CLAP}$_\text{src}\uparrow$}&\multicolumn{2}{c}{\textbf{FAD}$_\text{vgg}\downarrow$}
    \\
    \cline{2-5}
    &Full&Prefix&Full&Prefix
    \\
    \Xhline{3\arrayrulewidth}
    Drum-AIR&\textbf{0.619}&\textbf{0.627}&1.606&1.691\\
    Chord-AIR&0.614&0.625&1.593&1.681\\
    Piano-AIR&0.611&0.621&\textbf{1.531}&\textbf{1.623}\\
    \Xhline{2\arrayrulewidth}  
    MusicGen&0.373&0.441&2.474&2.276\\
    VampNet&0.613&0.618&3.689&3.910\\
    \Xhline{3\arrayrulewidth}
\end{tabular}
}
\caption{Evaluation of the models in terms of inpainting musicality. Adapter width $m=50$.}
\label{table:exp2}
\end{table}

\begin{table}[t]
\renewcommand{\arraystretch}{1.3}
  \setlength{
\tabcolsep}{2mm}
  \centering
  \subtable[Slakh2100 Test Set]{
  \begin{tabular}{ c|ccc }
    \Xhline{3\arrayrulewidth}
&\textbf{Drum}$_\text{SDR}\downarrow$&\textbf{Chord}$_\text{rec}\uparrow$&\textbf{Chroma}$_\text{cos}\uparrow$
      \\
\Xhline{2\arrayrulewidth}
Drum-AIR&6.676&-&-\\
Chord-AIR&-&0.745&0.830\\
Piano-AIR&-&0.720&0.849\\
\Xhline{3\arrayrulewidth}
  \end{tabular}}
  \subtable[RWC-POP-100]{
   \begin{tabular}{ c|ccc }
    \Xhline{3\arrayrulewidth}
&\textbf{Drum}$_\text{SDR}\downarrow$&\textbf{Chord}$_\text{rec}\uparrow$&\textbf{Chroma}$_\text{cos}\uparrow$
      \\
\Xhline{2\arrayrulewidth}
Drum-AIR&6.531&-&-\\
Chord-AIR&-&0.650&0.837\\
Piano-AIR&-&0.615&0.845\\
\Xhline{3\arrayrulewidth}  \end{tabular}}
  \caption{Evaluation of the models in terms of controllability. Adapter width $m=50$.}
  \label{table:exp1}
\end{table}

\begin{figure}[t]
  \centering
  \subfigure[Drum track controls, where the condition is drum audio.]
{\includegraphics[width=1\columnwidth]{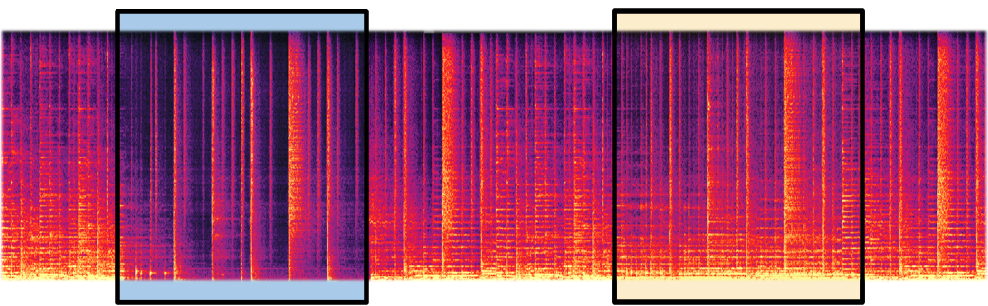}}
\subfigure[Chord progression controls, where the condition is block chords audio.]
{\includegraphics[width=1\columnwidth]{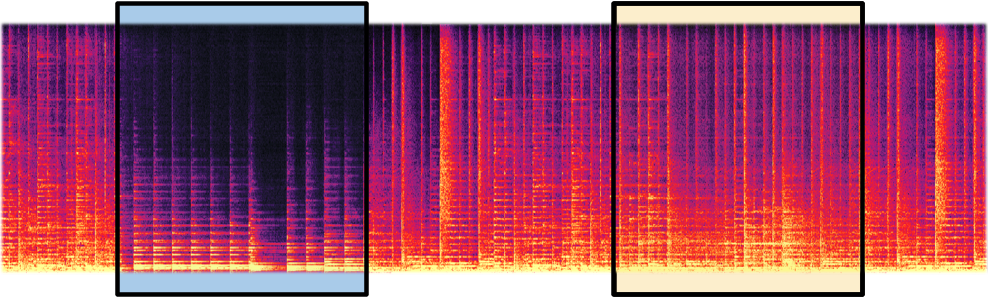}}
\subfigure[Arrangement and orchestration from piano cover, where the condition is piano cover audio.]
{\label{fig:spec1:3a}\includegraphics[width=1\columnwidth]{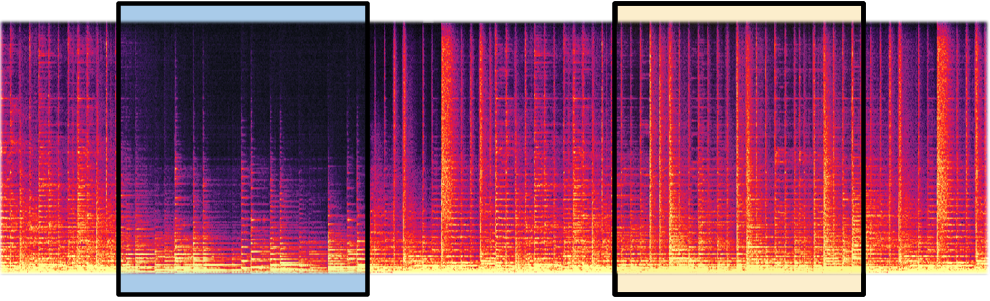}}
  \caption{Inpainting with various controls. The spectrogram conceptually repeats the same music segment twice, with the left half ($1,...,T$) providing the model with the infilling context and the controls, and the right half ($T+1,...,2T$) for the model to predict. Specifically, in the left half, outside the blue rectangle is the unmasked infilling context, and inside the blue rectangle is the separated/synthesized/user-specified audio representation of the condition (e.g., (a)~drum, (b)~block chords, (c)~piano cover). In the right half, the orange rectangle highlights the inpainted results.}
  \label{fig:spec1}
\end{figure}

Throughout the training process, we uniformly sample a mask ratio from the interval $[0.4, 0.8]$ and generate a binary mask with this ratio. To maintain the continuity of audio segments, we utilize a median filter with a window size of 11 to smooth the mask before applying it to the input Encodec tokens. 

We assess the performance of our models using two distinct datasets: the synthetic instrumental dataset (Slakh2100 Test Set) and a real-recording vocal dataset (RWC-POP-100). From each test song, we extract a 15-second audio segment for evaluation. We present three types of masks to corrupt the test audio, resulting in discontinuous masking areas: one, two, and four masking areas, respectively, referred to as Mask Pattern~1 through 3, as shown in Figure~\ref{fig:mask-123}. The total duration of masking areas sums up to 7.5s for each setting. 

The metrics we employ are as follows:

\begin{enumerate}
    \item \textbf{Source-to-Distortion Ratio (SDR)}. To evaluate how closely Drum-AIR adheres to the drum-track condition, we measure the SDR between the inpainted drum tracks (separated by Demucs) and the condition drum tracks. This measurement uses the latest SDR definition as specified in the Music Demixing Challenge (MDX 2021) competition \cite{mitsufuji2022music}.
    \item \textbf{Chord Accuracy}. For both Chord-AIR and Piano-AIR, we evaluate adherence to the condition chord progression by comparing the inpainted chords (predicted by a chord recognition model~\cite{jiang2019large}) with the condition chords. We report the weighted recall scores with a time resolution of 20ms to measure chord accuracy.
    \item \textbf{Chroma Cosine Similarity}. To assess how well Chord-AIR and Piano-AIR adhere to the chord and piano conditions, we calculate the chroma cosine similarity between the inpainted audios and the condition tracks, with a window size of 2048 and a hop size of 640 for the chromagram. 
    \item \textbf{CLAP Score}. For each model, we report the CLAP score \cite{wu2023large}, which assesses the semantic distance between the inpainted audio and the original audio.
    \item \textbf{Fréchet Audio Distance (FAD)}. For each model, we calculate the Frechet Audio Distance (FAD) of Visual Geometry Group (VGG) features between the inpainted audio and the original audio.
\end{enumerate}

\subsection{Baselines}
We designate MusicGen~\cite{copet2023simple} and VampNet~\cite{garcia2023vampnet} as our baselines, as there are currently no other comparable content-based controllable inpainting models available for comparison. As MusicGen can only perform continuation rather than inpainting, we only evaluate MusicGen using the past context of Mask Pattern~1.

\begin{enumerate}
    \item \textbf{MusicGen}. We reproduce the MusicGen model with the \textit{MusicGen-large} official checkpoint.\footnote{\url{https://huggingface.co/facebook/musicgen-large}.}
    \item \textbf{VampNet}. VampNet employs a non-autoregressive, bidirectional transformer with a variable masking schedule, enabling high-fidelity music synthesis across various tasks, such as inpainting and outpainting, through flexible inference masks. We reproduce the VampNet model with the the official code\footnote{\url{https://github.com/hugofloresgarcia/vampnet}.} and checkpoint\footnote{\url{https://zenodo.org/record/8136629/}.}. 
\end{enumerate}

\subsection{Results}
As illustrated in Table~\ref{table:exp2} and Table~\ref{table:exp1}, our model demonstrates competitive performance compared to unconditional generation and inpainting models. Furthermore, it showcases promising steerabilty.

\paragraph{Long-Gap Inpainting.} Table~\ref{table:exp2} presents the semantic similarity between inpainted audios and the original recordings, along with the audio qualities. In this scenario, MusicGen receives the prefix of Mask Pattern~1 (3.75s) and is tasked with generating an 11.25-second continuation. Conversely, VampNet and AIR models are conditioned with both the prefix and suffix of Mask Pattern~1. We present metrics for prefix-CLAP and prefix-FAD, evaluating the predicted (repeated) prefix and inpainted segments (11.25s), while full-CLAP and full-FAD assess both the predicted prefix, the inpainted segment, and the predicted suffix.

The comparison between prefix-CLAP and full-CLAP values in Table~\ref{table:exp2}(a) indicates that as MusicGen generates longer continuations, it ``strays further'' from the given context. When employing our heterogeneous adapter to enable inpainting settings for MusicGen, the resulting audio exhibits promising fidelity in preserving the original semantic context during generation. Furthermore, the low FAD further indicates that audio samples inpainted by our model exhibit competitive naturalness compared to those generated by the vallina MusicGen without suffix constraints. Compared to VampNet, our model achieves superior long-gap inpainting, as evidenced by the FAD value.

Table~\ref{table:exp2}(b) presents the results on the real-recording dataset with vocals, which is Out-Of-Distribution (OOD) w.r.t. the training dataset of MusicGen and our models but not w.r.t. VampNet. Comparing these results with those in Table~\ref{table:exp2}(a), it is evident that our models exhibit superior generalization ability, outperforming the vanilla MusicGen.

\begin{figure}[t]
  \centering
{\includegraphics[width=\columnwidth]{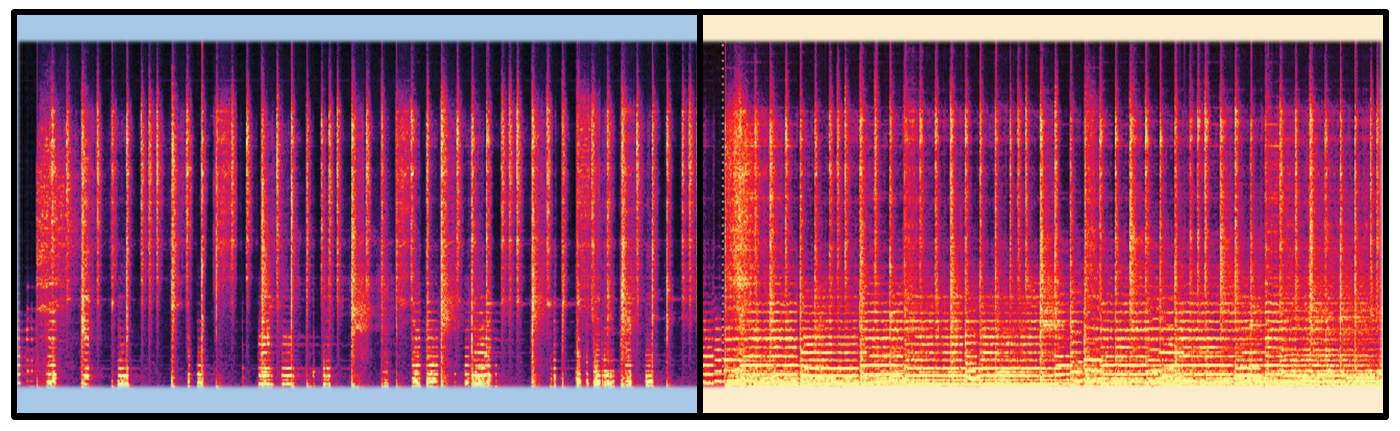}}
  \caption{
  Music audio inpainting under a masking rate of 0.7. In this example, 70\% of all frames are masked away and replaced by the drums condition. In the spectrogram above, the music audio can be identified by visible harmonics in the low-frequency region. Where those harmonics are absent there are the drum condition segments. Even under this setup, the model works well at the conditioned inpainting task.
}
  \label{fig:exp-mask70}
\end{figure} 

\begin{table*}[t]
\renewcommand{\arraystretch}{1.36}
  \setlength{
\tabcolsep}{2.2mm}
  \centering
  \begin{tabular}{ c|c|c|ccc|ccc }
    \Xhline{3\arrayrulewidth}
\multirow{ 2}{*}{$m$}&Trainable&Storage&
\multicolumn{3}{c|}{Slakh2100 Test Set}&\multicolumn{3}{c}{RWC-POP-100} \\
\cline{4-9}
&parameters&Space&\textbf{Drum}$_\text{SDR}\uparrow$&\textbf{CLAP}$_\text{src}\uparrow$&\textbf{FAD}$_\text{vgg}\downarrow$&\textbf{Drum}$_\text{SDR}\uparrow$&\textbf{CLAP}$_\text{src}\uparrow$&\textbf{FAD}$_\text{vgg}\downarrow$\\
\Xhline{2\arrayrulewidth}
10&3.75M&15M&\textbf{6.475}&0.758&\textbf{1.416}&6.109&\textbf{0.652}&\textbf{1.634}\\
30&11.25M&45M&6.257&\textbf{0.761}&1.434&6.479&0.646&1.705\\
50&18.75M&75M&6.628&0.759&1.442&\textbf{6.628}&0.643&1.673\\
\Xhline{3\arrayrulewidth}
  \end{tabular}
  \caption{The performance of models with different $m$ on Mask Pattern~2.}
  \label{table:exp3}
\end{table*}

\begin{figure}[t]
  \centering
  \subfigure[$g^l_1$]{\label{fig:clap-slakh}\includegraphics[width=0.48\columnwidth]{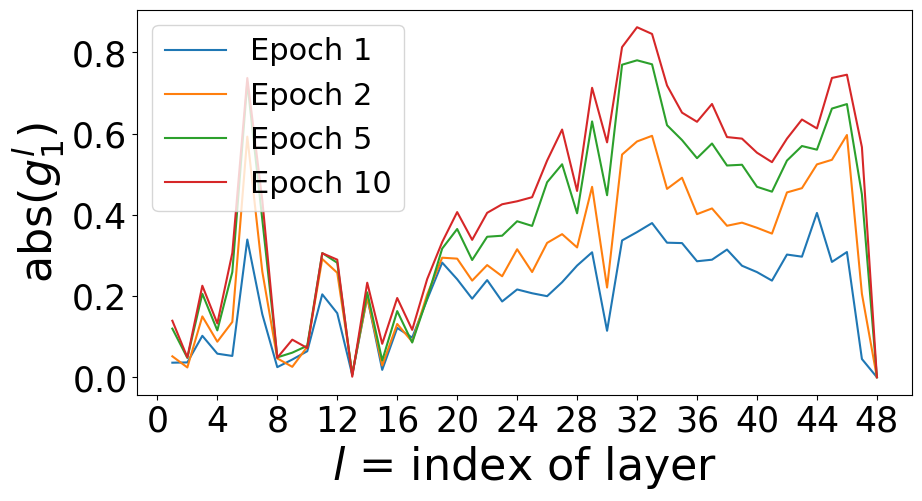}}
  \subfigure[$g^l_2$]{\label{fig:clap-rwc}\includegraphics[width=0.48\columnwidth]{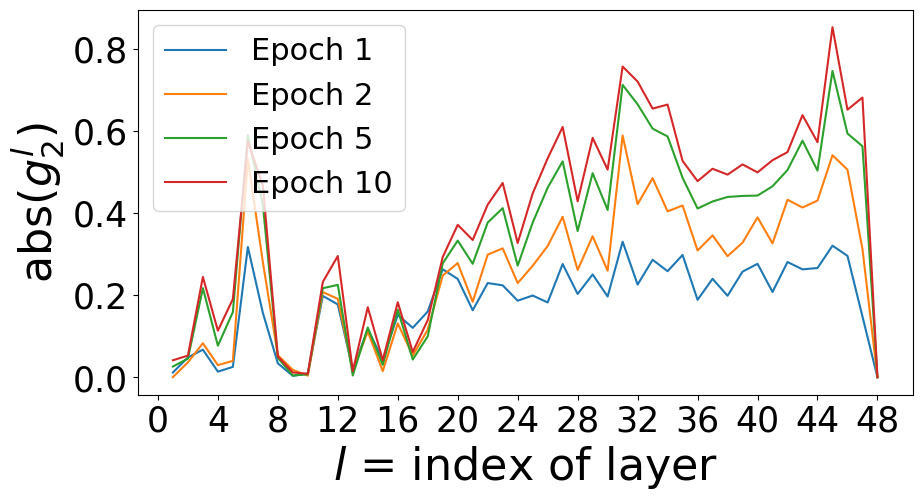}}
  \subfigure[$g^l_3$]{\label{fig:fad-slakh}\includegraphics[width=0.48\columnwidth]{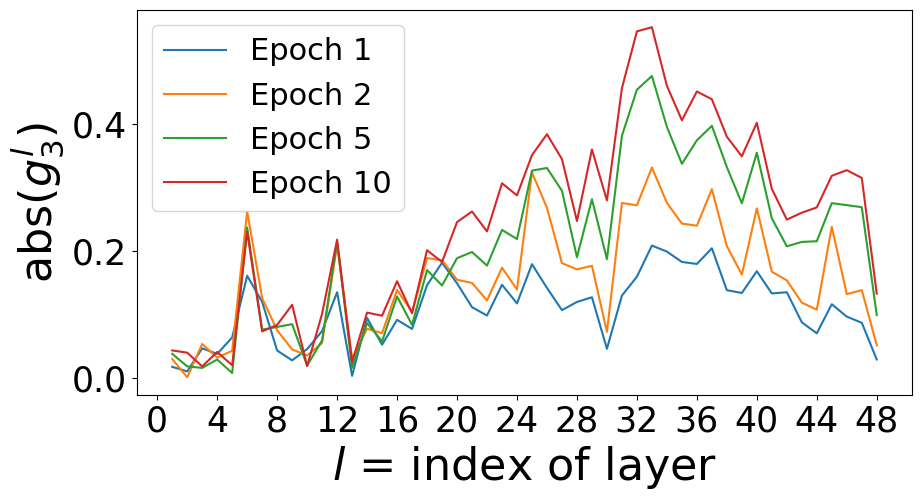}}
  \subfigure[$g^l_4$]{\label{fig:fad-rwc}\includegraphics[width=0.48\columnwidth]{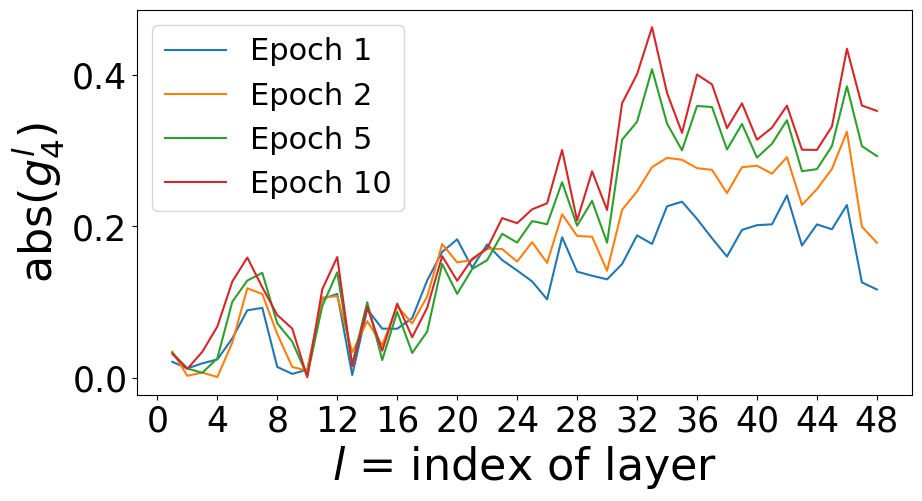}}
  \caption{The variation of $|g^l_i|$ across different types of adapters during training.}
  \label{fig:exp:gates}
\end{figure}

\paragraph{Controllable Inpainting.} As demonstrated in Table~\ref{fig:editing}, our model excels in controllable inpainting across various music editing tasks. The high SDR values in Table~\ref{table:exp1} indicate our model's proficiency in internal control, effectively inpainting the masked areas while preserving the condition track. Additionally, chord, chroma, and beat metrics underscore the capability of our models to integrate controls. Furthermore, CLAP values in Table~\ref{table:exp2} highlight that applying content-based controls extracted from the original audio serve as an auxiliary mechanism for maintaining the semantics of the original audio.


\paragraph{Arrangement and Refinement.} Figure~\ref{fig:spec1} illustrates examples of the results for the score-conditioned arrangement and the track-conditioned refinement tasks, highlighting our model's capability to reconstruct missing segments of the spectrogram. This demonstrates our approach's adaptability in performing spectrogram inpainting under various conditions.

Figure~\ref{fig:exp-mask70} presents a case of inpainting where 70\% of the spectrogram was intentionally masked prior to the inpainting process. Despite the significant loss of information, our model is able to restore the music composition, showing its robustness in handling substantial data gaps.

\paragraph{Low-Resource Fine-Tuning.} As shown in Table~\ref{table:exp3}, our heterogeneous adapters demonstrate a notable advantage in low-resource settings, with trainable parameters totaling less than 20M and a storage space requirement of less than 75M for the largest model. In comparison with the 3.3B parameters of vanilla MusicGen, our method effectively facilitates fine-tuning, converting an autoregressive music audio LLM into a steerable inpainting model.

\subsection{Ablation}
\label{sec:ablation}
\paragraph{Impact of Different Adapter Size.} A smaller $m$ implies a reduced size of adapters, resulting in fewer trainable parameters. However, Table~\ref{table:exp3} illustrates that a small $m$ does not significantly impair performance, highlighting the effectiveness of the proposed heterogeneous adapters. Additionally, Figure~\ref{fig:exp:gates} displays the variation of absolute values of learnable gates $g^l_{r(t)}$ in different layers during training. It reveals that the adapter influences deep layers more than shallow layers and that the adapters ${\bm a}^l_{\leq 2}$ for the prefix area encode more information than the prediction area ${\bm a}^l_{>2}$.

\paragraph{Impact of Different Mask Patterns.} During the ablation study, we utilize three different mask patterns, as illustrated in Figure~\ref{fig:mask-123}, and report the FAD values of inpainted audios in Table~\ref{table:exp4}. The context becomes more continuous, and the gap to be filled becomes shorter as we progress from Mask Pattern~1 to Mask Pattern~3. 
The results indicate that our model is highly robust to different mask structures and demonstrates a promising long-gap inpainting capability. Conversely, the baseline model's performance deteriorates as the masked gap lengthens.

\begin{table}[t]
\renewcommand{\arraystretch}{1.3}
  \setlength{
\tabcolsep}{3mm}
  \centering
  \subtable[Slakh2100 Test Set]{
  \begin{tabular}{ c|ccc }
    \Xhline{3\arrayrulewidth}
&Mask~1&Mask~2&Mask~3
\\
\Xhline{3\arrayrulewidth}
Drum-AIR&1.423&1.442&1.467\\
VampNet&2.910&2.390&2.091\\
\Xhline{3\arrayrulewidth}
  \end{tabular}}
  \subtable[RWC-POP-100]{
  \begin{tabular}{ c|ccc }
    \Xhline{3\arrayrulewidth}
&Mask~1&Mask~2&Mask~3
\\
\Xhline{3\arrayrulewidth}
Drum-AIR&1.606&1.673&1.759\\
VampNet&3.689&3.496&3.336\\
\Xhline{3\arrayrulewidth}
  \end{tabular}}
  \caption{The FAD values of generated audio with different mask patterns. A lower value indicates higher audio quality.}
  \label{table:exp4}
\end{table}

\section{Conclusions}
In this paper, we propose a novel Parameter-Efficient Fine-Tuning (PEFT) method and apply it to MusicGen. It enables the model to inpaint arbitrary segments of music, while retaining the generation quality of MusicGen. Furthermore, we introduce frame-level content-based controls during fine-tuning, and demonstrate it with conditional generation tasks given the drum track, chord sequence, or piano cover. The model shows improved performance over baseline methods on several metrics. The result shows strong usability and flexibility for future music editing tools.

There are several potential improvements for the method. First, the unmasked infilling context weakens the effectiveness of the text prompt for MusicGen. A different fine-tuning setting might be required to resolve the problem. Other future works include introducing more content-based controls, as well as conditional generation based on multiple content-based controls simultaneously.

\appendix

\section*{Acknowledgments}
This work was supported in part through the NYU and NYUSH IT High Performance Computing resources, services, and staff expertise.

\bibliographystyle{named}
\bibliography{ijcai24}

\end{document}